\documentclass[aps,prb,twocolumn]{revtex4-1}
\newcommand{\bdot}{\mbox{\boldmath $\,\cdot\,$}}
\newcommand{\bnabla}{\mbox{\boldmath $\nabla$}}

\usepackage{graphicx}
\usepackage{amssymb}
\usepackage{amsmath}
\usepackage{cancel}

\def\ihat{{\bf \hat x}}
\def\jhat{{\bf \hat y}}
\def\khat{{\bf \hat z}}

\def\rhat{{\bf \hat r}}

\begin{document}

\title{Mansuripur's paradox}

\author{David J.~Griffiths}\email[Electronic address: ]{griffith@reed.edu}
\affiliation{Department of Physics, Reed College, Portland, Oregon  97202}
\author{V.~Hnizdo}
\affiliation{National Institute for Occupational Safety and Health, Morgantown, West Virginia 26505}

\begin{abstract}
A recent article claims that the Lorentz force law is incompatible with special relativity.  We discuss the ``paradox" on which this claim is based.  The resolution depends on whether one assumes a ``Gilbert" model for the magnetic dipole (separated monopoles) or the standard ``Amp\`ere" model (a current loop).  The former case was treated in these pages many years ago; the latter, as several authors have noted, constitutes an interesting manifestation of ``hidden momentum."
\end{abstract}

\maketitle

\section{Introduction}
On May 7, 2012, a remarkable article appeared in {\it Physical Review Letters}. \cite{Man}  The author, Masud Mansuripur, claimed to offer ``incontrovertible theoretical evidence of the incompatibility of the Lorentz [force] law with the fundamental tenets of special relativity," and concluded that ``the Lorentz law must be abandoned."  The Lorentz law,
\begin{equation}
{\bf F}= q[{\bf E} + ({\bf v} \times {\bf B})]
\end{equation}
tells us the force {\bf F} on a charge $q$ moving with velocity {\bf v} through electric and magnetic fields {\bf E} and {\bf B}. Together with Maxwell's equations, it is the foundation on which all of classical electrodynamics rests.  If it is incorrect, 150 years of theoretical physics is in serious jeopardy.

Such a provocative proposal was bound to attract attention.  {\it Science} \cite{SCI} published a full-page commentary,
and within days several rebuttals were posted. \cite{Reb}  Critics pointed out that since the Lorentz force law can be embedded in a manifestly covariant formulation of electrodynamics, it is guaranteed to be consistent with special relativity,\cite{COV} and   and some of them identified the specific source of Mansuripur's error: neglect of ``hidden momentum."  Nearly a year later {\it Physical Review Letters} published four rebuttals, \cite{YL} and {\it Science} printed a follow-up article declaring the ``purported relativity paradox resolved." \cite{SCI2}

Mansuripur's argument is based on a ``paradox" that was explored in this journal by Victor Namias and others \cite{Namias} many years ago: a magnetic dipole moving through an electric field can experience a torque, with no accompanying rotation.  In Section II we introduce Mansuripur's version of the paradox, in simplified form, and explain Namias's resolution.  The latter is based on a ``Gilbert" model of the dipole (separated magnetic monopoles); it does not work for the (realistic) ``Amp\`ere" model (a current loop).  For Amperian dipoles the resolution involves ``hidden" momentum, so in Section III we discuss the physical nature of this often-misunderstood phenomenon.  Mansuripur himself treated the dipole as the point limit of a magnetized object, so in Section IV we repeat the calculations in that context (for both models), and confirm our earlier results.  In Section V we discuss the Einstein--Laub force law, which Mansuripur proposed as a replacement for the Lorentz law, and in Section VI we offer some comments and conclusions.

\section{Gilbert Dipoles: Namias's Resolution}

\vskip0in
\begin{figure}[b]
\hskip-.2in\scalebox{.7}[.7]{\includegraphics{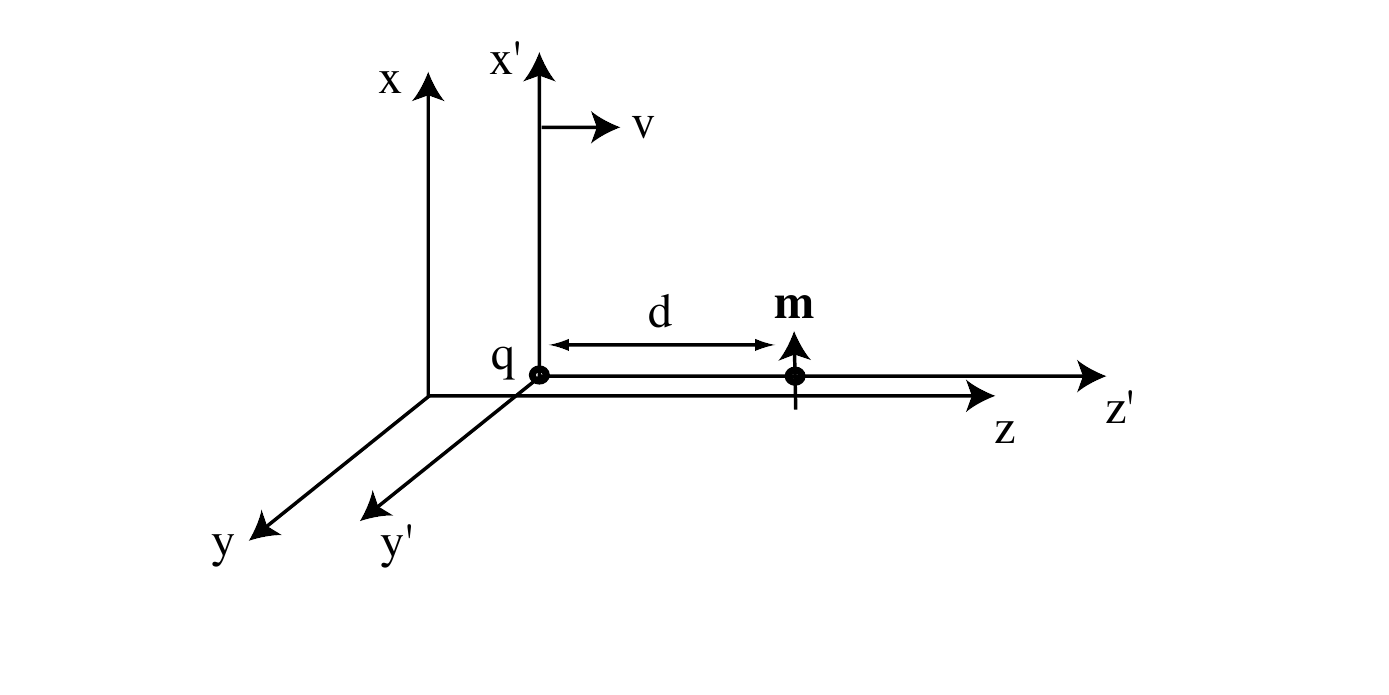}}
\vskip-.3in
\caption{Electric charge ($q$) and magnetic dipole ({\bf m}) in proper (primed) and lab (unprimed) frames.}
\end{figure}

First the paradox:  In ${\cal S}'$ (the ``proper" frame) there is an ideal magnetic dipole ${\bf m} = m_0\,\ihat$ at $(0,0,d)$, and a point charge $q$ at the origin, both at rest.  The torque on {\bf m} is (obviously) zero.  Now examine the same configuration in ${\cal S}$ (the ``lab" frame), with respect to which ${\cal S}'$ moves at constant speed $v$ in the $z$ direction (Fig.~1).  In ${\cal S}$ the (moving) point charge generates electric and magnetic fields
\begin{eqnarray}
{\bf E}(x,y,z,t) &= &\frac{q}{4\pi\epsilon_0}\frac{\gamma}{R^3}\left(x\,\ihat + y\,\jhat + (z-vt)\,\khat\right),\\
{\bf B}(x,y,z,t) &= &\frac{q}{4\pi\epsilon_0}\frac{v\gamma}{c^2R^3}\left(-y\,\ihat + x\,\jhat\right),
\end{eqnarray}
($\gamma \equiv 1/\sqrt{1-(v/c)^2}$, $R\equiv \sqrt{x^2+y^2+\gamma^2(z-vt)^2}\,$), and the (moving) magnetic dipole acquires an electric dipole moment \cite{VH}
\begin{equation}
{\bf p} = \frac{1}{c^2}({\bf v}\times {\bf m})=\frac{1}{c^2}vm_0\,\jhat.
\end{equation}

\noindent The torque on the dipole is
\begin{equation}
{\bf N} = ({\bf m}\times {\bf B}) + ({\bf p}\times {\bf E})  = \frac{qm_0}{4\pi\epsilon_0}\frac{v}{c^2d^2}\,\ihat
\end{equation}
(by Lorentz transformation, $d = \gamma(z-vt)$; the magnetic contribution is zero, because {\bf B} vanishes on the $z$ axis).  The torque is zero in one inertial frame, but {\it non}-zero in the other!  Mansuripur concludes that the Lorentz force law (on which Eq.~5 is predicated) is inconsistent with special relativity.

\vskip0in
\begin{figure}[t]
\hskip-.2in\scalebox{.7}[.7]{\includegraphics{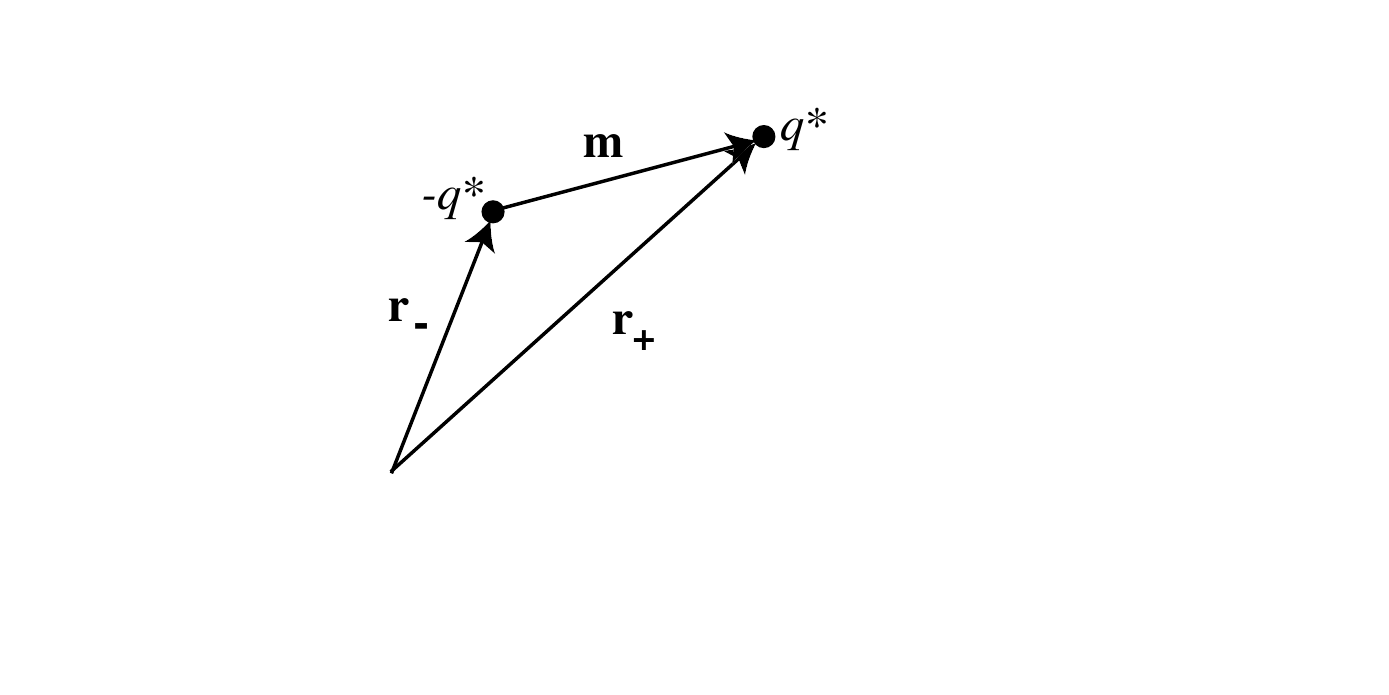}}
\vskip-.7in
\caption{``Gilbert" magnetic dipole.}
\end{figure}

This ``paradox" was resolved years ago by Victor Namias. \cite{Namias}  The standard torque formulas (${\bf p}\times {\bf E}$ and ${\bf m}\times {\bf B}$) apply to dipoles {\it at rest}, but they do not hold, in general, for dipoles in motion.  Suppose we model the magnetic dipole as separated monopoles (Fig. 2).   The ``Lorentz force law" for a magnetic monopole $q^*$ reads \cite{DJG1}
\begin{equation}
{\bf F} = q^*\left[{\bf B} - (1/c^2){\bf v}\times {\bf E}\right],
\end{equation}
so the torque \cite{origin} on a moving dipole ${\bf m}=q^*({\bf r}_+-{\bf r}_-)$ is
\begin{eqnarray*}
{\bf N} &=& ({\bf r}_+\times {\bf F}_+)+({\bf r}_-\times {\bf F}_-)\\
&=&({\bf m}\times {\bf B})-\frac{1}{c^2}{\bf m}\times({\bf v}\times {\bf E}).
\end{eqnarray*}
But ${\bf m}\times({\bf v}\times {\bf E})= {\bf v}\times({\bf m}\times {\bf E})+({\bf m}\times{\bf v})\times {\bf E}$, so
\begin{eqnarray}
{\bf N} &=& ({\bf m}\times {\bf B})-\frac{1}{c^2}({\bf m}\times{\bf v})\times {\bf E} - \frac{1}{c^2}{\bf v}\times({\bf m}\times{\bf E})\nonumber\\
&=& ({\bf m}\times {\bf B})+({\bf p}\times {\bf E}) - \frac{1}{c^2}{\bf v}\times({\bf m}\times{\bf E}).
\end{eqnarray}
There is a third term, missing in Eq.~5, which (it is easy to check) exactly cancels the offending torque; the net torque is zero in both frames.

\section{Amp\`ere Dipoles: Hidden Momentum}

Namias believed that his formula (Eq.~7) applies just as well to an Amp\`ere dipole as it does to a Gilbert dipole.   He was mistaken.  An Amp\`ere dipole in an electric field carries ``hidden" momentum, \cite{hid_mom}
\begin{equation}
{\bf p}_h = \frac{1}{c^2}({\bf m}\times {\bf E}).
\end{equation}
Because it is crucial in understanding the resolution to Mansuripur's paradox, we pause to review the derivation of this formula, in a simple model.

Imagine a rectangular loop of wire carrying a steady current.  Picture the current as a stream of noninteracting positive charges that move freely within the wire. \cite{UN}  When a uniform electric field {\bf E} is applied (Fig.~3), the charges accelerate up the left segment, and decelerate down the right one.  {\it Question:} What is the total momentum of all the charges in the loop?
The left and right segments cancel, so we need only consider the top and bottom.  Say there are $N_t$ charges in the top segment, going to the right at speed $v_t$, and $N_b$ charges in the lower segment, going at (slower) speed $v_b$ to the left.  The {\it current} ($I=\lambda v$) is the same in all four segments (otherwise charge would be piling up somewhere).  Thus
\begin{equation}
I = {qN_t\over l}v_t= {qN_b\over l}v_b,\quad {\rm so}\quad N_tv_t= N_bv_b= {Il\over q},
\end{equation}
where $q$ is the charge of each particle, and $l$ is the length of the rectangle.  {\it Classically,} the momentum of a single particle is $p=mv$, where $m$ is its mass, so the total momentum (to the right) is
\begin{equation}
p_{\rm classical} = mN_tv_t-mN_bv_b = m{Il\over q}-m{Il\over q} = 0,
\end{equation}
as one would certainly expect (after all, the loop as a whole is not moving).  But {\it relativistically} the momentum of a particle is $p=\gamma m v$, and we get
\begin{equation}
p_{\rm relativistic} = \gamma_tmN_t v_t-\gamma_bmN_b v_b= {mIl\over q}\left(\gamma_t-\gamma_b\right),
\end{equation}
which is {\it not} zero, because the particles in the upper segment are moving faster.
In fact, the gain in energy ($\gamma m c^2$), as a particle goes up the left side, is equal to the work done by the electric force, $qEw$, where $w$ is the height of the rectangle, so
\begin{equation}
\gamma_t-\gamma_b = {qEw\over mc^2},\quad {\rm and\ hence}\quad p_{\rm rel} = {IlEw\over c^2}.
\end{equation}
Now $Ilw$ is the magnetic dipole moment of the loop; as vectors, {\bf m} points into the page and {\bf p} is to the right, so
\begin{equation}
{\bf p}_{\rm rel} = {1\over c^2}({\bf m}\times {\bf E}).
\end{equation}
This is the ``hidden" momentum in Eq.~8.

\vskip0in
\begin{figure}[t]
\hskip-1.05in\scalebox{.8}[.8]{\includegraphics{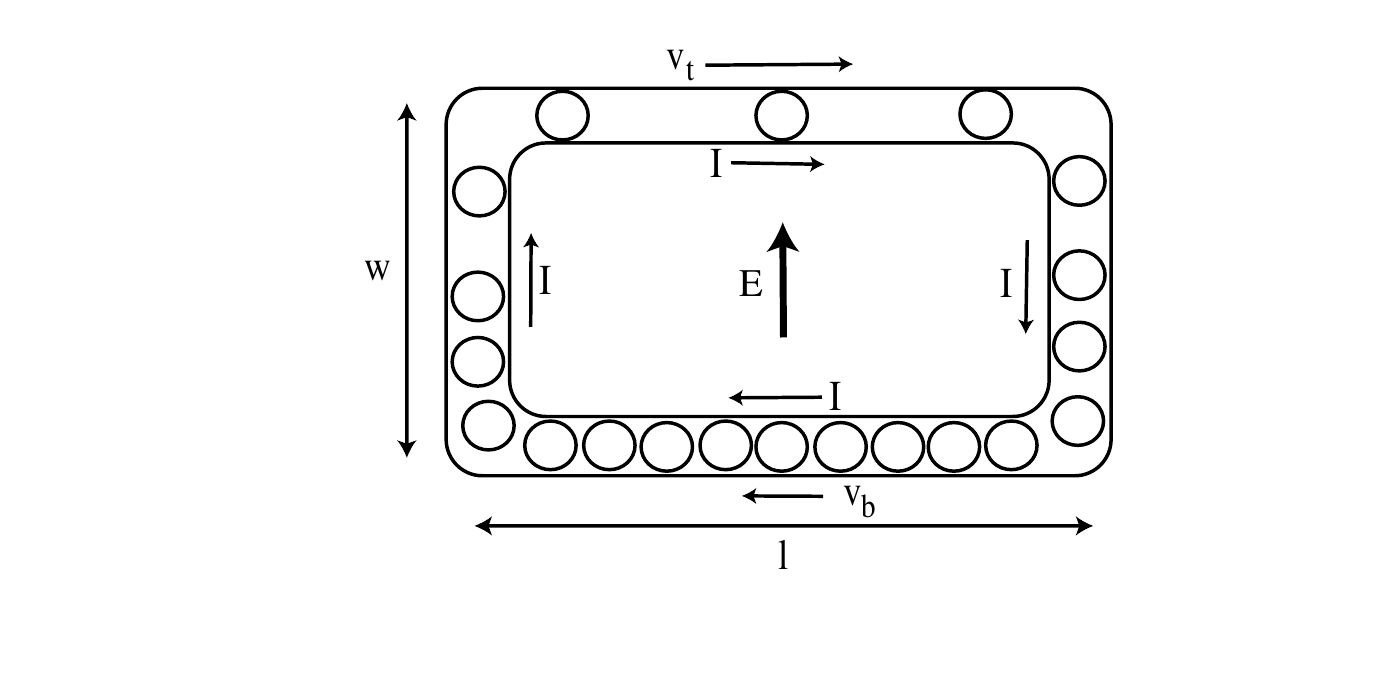}}
\vskip-.5in
\caption{Current loop in an external electric field.}
\end{figure}

The term ``hidden momentum" was coined by Shockley; \cite{hid_mom} it was an unfortunate choice.  The phenomenon itself was first studied in the context of static electromagnetic systems with nonzero field momentum (${\bf p}_{\rm field} = \epsilon_0\int({\bf E}\times{\bf B})\,d^3{\bf r}$).  In such configurations the hidden momentum exactly cancels the field momentum (${\bf p}_h=-{\bf p}_{\rm field}$), leaving a total of zero, as required by the ``center of energy theorem." \cite{CET} This has created the impression that hidden momentum is something artificial and ad hoc---invented simply to rescue an abstract theorem. \cite{ABS} \, Nothing could be farther from the truth.  Hidden momentum is perfectly ordinary relativistic mechanical momentum, as the example above indicates; it occurs in systems with internally moving parts, such as current-carrying loops, and it is ``hidden" only in the sense that it not associated with motion of the object as a whole.   A {\it Gilbert} dipole in an electric field, having no moving parts, harbors {\it no} hidden momentum (and the fields---with the crucial delta-function term in {\bf B} included---carry no compensating momentum). \cite{HM}

Returning to the configuration in Fig.~1, the hidden momentum in ${\cal S}'$ is
\begin{equation}
{\bf p}_h=\frac{1}{c^2}\left[(m_0\,\ihat) \times \left(\frac{1}{4\pi\epsilon_0}\frac{q}{d^2}\,\khat\right)\right] = -\frac{qm_0}{4\pi\epsilon_0 c^2d^2}\,\jhat.
\end{equation}
Because ${\bf p}_h$ is perpendicular to {\bf v}, and transverse components are unaffected by Lorentz transformations, this is also the hidden momentum in $\cal S$.  It is constant (in time), so there is no associated force.   But the hidden {\it angular} momentum,
\begin{equation}
{\bf L}_h = {\bf r}\times{\bf p}_h,
\end{equation}
is {\it not} constant (in the lab frame), because {\bf r} is changing.  In fact,
\begin{equation}
\frac{d{\bf L}_h}{dt} = {\bf v} \times {\bf p}_h = \frac{qm_0}{4\pi\epsilon_0}\frac{v}{c^2d^2}\,\ihat.
\end{equation}
This increase in angular momentum requires a torque,
\begin{equation}
{\bf N} = \frac{qm_0}{4\pi\epsilon_0}\frac{v}{c^2d^2}\,\ihat,
\end{equation}
and this is precisely what we found in Eq.~5.

{\it Recapitulating:} In the Gilbert model there is an extra term in the torque formula (Eq.~7); the total torque is zero, there is no hidden angular momentum, and nothing rotates.  In the Amp\`ere model there is no third term in the torque formula (Eq.~5)\cite{NTT}; the torque is {\it not} zero, and drives the increasing hidden angular momentum---but still nothing rotates.\cite{lab_torque}  It helps to separate the angular momentum into two types: ``overt" (associated with actual rotation) and ``hidden" (so called because it is {\it not} associated with any overt rotation of the object).  Torque is the rate of change of the {\it total} angular momentum:
\begin{equation}
{\bf N} =  \frac{d{\bf L}_o}{dt}+\frac{d{\bf L}_h}{dt}.
\end{equation}
In both models $d{\bf L}_o/dt={\bf 0}$.  In the Gilbert model {\bf N} and $d{\bf L}_h/dt$ are also zero; in the Amp\`ere model they are equal but non-zero.

\section{Magnetized Materials}

It is of interest to see how this plays out in Mansuripur's formulation of the problem.  He treats the dipole as magnetized medium, and calculates the torque directly from the Lorentz force law, without invoking ${\bf p}\times {\bf E}$ or ${\bf m}\times {\bf B}$.  In the proper frame, he takes
\begin{equation}
{\bf M}'(x',y',z',t')= m_0\delta(x')\delta(y')\delta(z'-d)\,\ihat.
\end{equation}
Now, {\bf M} and {\bf P} constitute an antisymmetric second-rank tensor:
\begin{equation}
P^{\mu\nu} = \begin{pmatrix}0&cP_x&cP_y&cP_z\\-cP_x&0&-M_z&M_y\\-cP_y&M_z&0&-M_x\\-cP_z&-M_y&M_x&0,
\end{pmatrix}
\end{equation}
and the transformation rule is \cite{DJG2}
\begin{eqnarray*}
&&P_z=P_z',\ P_x=\gamma(P_x'+\frac{v}{c^2}M_y'),\ P_y=\gamma(P_y'-\frac{v}{c^2}M_x')\\
&&M_z=M_z',\ M_x=\gamma(M_x'-vP_y'),\ M_y=\gamma(M_y'+vP_x')
\end{eqnarray*}
(for motion in the $z$ direction).  In the present case, then, the polarization and magnetization in the ``lab" frame are
\begin{eqnarray}
{\bf M}(x,y,z,t)&= &m_0\delta(x)\delta(y)\delta\left(z-vt-(d/\gamma)\right)\ihat,\\
{\bf P}(x,y,z,t)&=& \frac{m_0v}{c^2}\delta(x)\delta(y)\delta\left(z-vt-(d/\gamma)\right)\jhat.
\end{eqnarray}

According to the Lorentz law, the force density is
\begin{equation}
{\bf f} = \rho {\bf E} + {\bf J}\times{\bf B},
\end{equation}
where $\rho = -\bnabla\bdot {\bf P}$ is the bound charge density and ${\bf J} = \partial {\bf P}/\partial t + \bnabla \times{\bf M}$ is the sum of the polarization current and the bound current density.  Using Eqs.~2, 3, 21, and 22, we obtain
\begin{eqnarray}
{\bf f} &=& -(\bnabla\bdot {\bf P}){\bf E} +(\bnabla \times {\bf M})\times {\bf B} + \frac{\partial {\bf P}}{\partial t}\times{\bf B}\nonumber\\
&=&-\frac{qm_0v}{4\pi\epsilon_0 c^2}\frac{d}{R^3}\delta(x)\delta'(y)\delta(z-vt-d/\gamma)\,\khat
\end{eqnarray}
(where a prime denotes the derivative).
The net force on the dipole is
\begin{equation}
{\bf F} = \int{\bf f}\,dx\,dy\,dz = \frac{qm_0vd}{4\pi\epsilon_0 c^2}\frac{d}{dy}\left[\frac{1}{(y^2+d^2)^{3/2}}\right]\Bigg|_{y=0}\,\khat =  {\bf 0}.
\end{equation}
Meanwhile, the torque density is
\begin{equation}
{\bf n} = {\bf r}\times {\bf f} = -\frac{qm_0vd}{4\pi\epsilon_0 c^2}\frac{y}{R^2}\delta(x)\delta'(y)\delta(z-vt-d/\gamma)\,\ihat,
\end{equation}
so the net torque on the dipole is
\begin{eqnarray}
{\bf N} &=& \int {\bf n}\,dx\,dy\,dz\nonumber\\
&=&-\frac{qm_0vd}{4\pi\epsilon_0 c^2}\left\{-\frac{d}{dy}\left[\frac{y}{(y^2+d^2)^{3/2}}\right]\right\}\Bigg|_{y=0}\,\ihat\nonumber\\
&=&\frac{qm_0v}{4\pi\epsilon_0 c^2d^2}\,\ihat,
\end{eqnarray}
confirming Eq.~5.  This is the torque required to account for the increase in hidden angular momentum.

What if we run Mansuripur's calculation for a dipole made out of magnetic monopoles?  The bound charge, bound current, and magnetization current are \cite{minus}
\begin{equation}
\rho^*_b = -\bnabla\bdot {\bf M},\quad {\bf J}^*_b = -c^2\bnabla\times {\bf P},\quad {\bf J}^*_p=\frac{\partial{\bf M}}{\partial t},
\end{equation}
so the force density on the magnetic dipole (again invoking Eqs.~2, 3, 21, and 22) is\cite{LFL}
\begin{eqnarray}
{\bf f}&=& \rho^*{\bf B} -\frac{1}{c^2}{\bf J}^*\times{\bf E} \nonumber\\
&=& -(\bnabla \bdot {\bf M}){\bf B}+(\bnabla \times{\bf P})\times{\bf E} - \frac{1}{c^2}\left(\frac{\partial {\bf M}}{\partial t}\right)\times {\bf E}\nonumber\\
&=& {\bf 0}.
\end{eqnarray}
The total force is again zero, but this time so too is the torque density (${\bf n} = {\bf r}\times {\bf f}$), and hence the total torque.  As before, the torque is zero in the Gilbert model---and there is no hidden angular momentum.

\section{The Einstein--Laub Force Law}

Having concluded that the Lorentz force law is unacceptable, Mansuripur proposes to replace Eq.~24 with an expression based on the Einstein--Laub law: \cite{EL}
\begin{eqnarray}
{\bf f}_{\rm EL}&=&({\bf P}\bdot\bnabla){\bf E}+\frac{\partial\bf P}{\partial t}\times(\mu_0{\bf H})
 +({\bf M}\bdot\bnabla) \mu_0{\bf H}\nonumber\\
 &&-\ \frac{1}{c^2}\,\frac{\partial\bf M}{\partial t}\times{\bf E}\nonumber\\
 &=&\frac{m_0qv\gamma}{4\pi\epsilon_0 c^2}\frac{1}{R^3}\delta(x)\delta(y)[2\delta(z-vt-d/\gamma)\nonumber\\
 &&-\,(z-vt)\delta'(z-vt-d/\gamma)]\,\jhat.
\end{eqnarray}
The total force on the dipole still vanishes:
\begin{eqnarray}
{\bf F}_{\rm EL}&=&
\frac{m_0qv\gamma}{4\pi\epsilon_0 c^2}\,\jhat\left\{\frac{2}{d^3} + \frac{1}{\gamma^3}\frac{d}{dz}\left[\frac{1}{(z-vt)^2}\right]\Bigg|_{z-vt=d/\gamma}\right\}\nonumber\\
&=& {\bf 0}.
\end{eqnarray}

The torque density should be ${\bf r}\times{\bf f}_{\rm EL}$:
\begin{eqnarray}
{\bf n}_{\rm EL}&=&
-\frac{m_0qv\gamma}{4\pi\epsilon_0 c^2}\frac{z}{R^3}\delta(x)\delta(y)[2\delta(z-vt-d/\gamma)\nonumber\\
&&\quad-\,(z-vt)\delta'(z-vt-d/\gamma)]\,\ihat,
\end{eqnarray}
giving a total torque
\begin{eqnarray}
{\bf N}_{\rm EL}
&=& -\,\frac{m_0qv\gamma}{4\pi\epsilon_0 c^2}\,\ihat\left\{\frac{2(vt+d/\gamma)}{d^3}+\frac{1}{\gamma^3}\frac{d}{dz}\left[\frac{z}{(z-vt)^2}\right]\right\}\nonumber\\
&=&-\frac{m_0qv}{4\pi\epsilon_0 c^2 d^2}\,\ihat
\end{eqnarray}
(the derivative is again evaluated at $z-vt=d/\gamma$).
It's not zero!  In fact, it's {\it minus} the ``Lorentz" torque, Eq.~27.  But Mansuripur argues that, ``To guarantee the conservation of angular momentum, [Eq.~32] must be supplemented \ldots"
\begin{equation}
{\bf n}_{\rm EL}' = {\bf n}_{\rm EL}  +({\bf P}\times{\bf E})+({\bf M}\times{\bf B}).
\end{equation}
In our case the extra terms are
\begin{equation*}
({\bf P}\times{\bf E})+({\bf M}\times{\bf B})=\frac{m_0qv}{4\pi\epsilon_0 c^2d^2}\delta(x)\delta(y)\delta(z-vt-d/\gamma)\,\ihat,
\end{equation*}
and their contribution to the total torque is
\begin{equation}
\int [({\bf P}\times{\bf E})+({\bf M}\times{\bf B})]\,dx\,dy\,dz = \frac{m_0qv}{4\pi\epsilon_0 c^2d^2}\,\ihat,
\end{equation}
which is just right to cancel Eq.~33, yielding a net torque of zero (which Mansuripur takes to be the correct answer).

What are we to make of this argument?  In the first place, the Einstein--Laub force density was derived assuming that the medium is at rest,\cite{EL} which in this case it is not.  More important, the magnetization terms implicitly assume a Gilbert model for the magnetic dipole:
\begin{equation}
 ({\bf M}\bdot\bnabla){\bf B}
 =-\bnabla\times({\bf M}\times{\bf B})+({\bf B}\bdot\bnabla){\bf M}-(\bnabla\bdot{\bf M}){\bf B};
 \end{equation}
as long as the magnetization is localized,  the first two terms yield vanishing surface integrals, \cite{SUI} leaving $-(\bnabla\bdot{\bf M}){\bf B}-(1/c^2)[(\partial {\bf M}/\partial t)\times {\bf E}]$ for the net force density on the object, the same as in the Gilbert model (Eq.~29). \cite{Tellegen}  There may be some contexts in which the Einstein--Laub force law is valid and useful, but this is not one of them.  Mansuripur is quite explicit in writing that the magnetic dipole he has in mind is ``a small, charge neutral loop of current," which is to say, an Amp\`ere dipole.

\section{Conclusion}

The resolution of Mansuripur's ``paradox" depends on the model for the magnetic dipole:
\begin{itemize}
\item If it is a Gilbert dipole (made from magnetic monopoles), the third term in Namias's formula (Eq.~7) supplies the missing torque.  In Mansuripur's formulation (using a polarizable medium), it comes from a correct accounting of the bound charge/current (Eq.~28).  The net torque is zero in the lab frame, just as it is in the proper frame.
\item If it is an Amp\`ere dipole (an electric current loop), the third term in Namias's equation is absent, and the torque on the dipole is {\it not} zero.  It is, however, just right to account for the increasing hidden angular momentum in the dipole.
\end{itemize}
In either model the Lorentz force law is entirely consistent with special relativity.

\bigskip

We thank Kirk McDonald, Daniel Vanzella, and Daniel Cross for useful correspondence.   VH coauthored this paper in his private capacity; no official support or endorsement by the Centers for Disease Control and Prevention is intended or should be inferred.


\begin{thebibliography}{99}
\bibitem{COV} The details are worked out in Cross and Vanzella (ref.~3).
\bibitem{UN} This is of course an unrealistic model for an actual current-carrying wire.  Vaidman (Ref.~11) explores more plausible models, but the result is unchanged.
\bibitem{NTT} We do not know a simple way to prove this directly, but we will confirm it implicitly in Section IV.
\end{thebibliography}
\end{document}